\documentclass[pra,twocolumn,showpacs,nofootinbib]{revtex4-1}

\usepackage{bm,dcolumn,amsmath,graphicx}
\usepackage{physdefs}

\def\cjp{Can. J. Phys.}

\def\epl{Europhys. Lett.}

\def\hypint{Hyp. Int.}

\def\mnras{Mon. Not. R. Astron. Soc.}

\def\pra{Phys. Rev. A}

\def\prl{Phys. Rev. Lett.}

\def\pscr{Phys. Scr.}

\def\rmp{Rev. Mod. Phys.}
\def\sci{Science}

\begin{document}
\title{Enhanced laboratory sensitivity to variation of the fine-structure constant using highly-charged ions}

\author{J. C. Berengut}
\author{V. A. Dzuba}
\author{V. V. Flambaum}
\affiliation{School of Physics, University of New South Wales, Sydney, NSW 2052, Australia}

\date{8 July 2010}

\pacs{06.30.Ft,37.10.Ty,31.15.am,32.30.Jc}
\begin{abstract}

We study atomic systems that are in the frequency range of optical atomic clocks and have enhanced sensitivity to potential time-variation of the fine structure constant $\alpha$. The high sensitivity is due to coherent contributions from three factors: high nuclear charge $Z$, high ionization degree, and significant differences in the configuration composition of the states involved. Configuration crossing keeps the frequencies in the optical range despite the large ionization energies. We discuss a few promising examples that have the largest $\alpha$-sensitivities seen in atomic systems.

\end{abstract}

\maketitle

\section{Introduction}

Theories that seek to unify gravity with the other fundamental interactions suggest that temporal variation of fundamental constants is a possibility, or even a necessity, in an expanding Universe (see, e.g.~\cite{uzan03rmp}). Hints from quasar absorption spectra that the fine-structure constant, $\alpha = e^2/\hbar c$, may have been different in the distant past have been reported~\cite{webb99prl,murphy03mnras} but have not been confirmed by other groups working on a different telescope~\cite{srianand04prl}. More recently, the methodology of~\cite{srianand04prl} was questioned~\cite{murphy07prl} and a reanalysis of the same data increased the reported error bars by a factor of six~\cite{murphy08mnras}. The observational status is therefore still unclear.

Atomic clocks provide a complementary method to search for temporal variation of fundamental constants in terrestrial laboratories~\cite{dzuba99prl,berengut10hypint}. Two different clocks are compared over the course of several years and any discrepancies are interpreted in terms of variation of $\alpha$. For such a scheme it is critical that the two clocks have different sensitivities to any possible variation of $\alpha$. The best current laboratory limit, $\dot\alpha/\alpha = (-1.6\pm2.3)\E{-17}$~year$^{-1}$, comes from comparison of Hg$^+$ and Al$^+$ optical clocks over the course of a year~\cite{rosenband08sci}. Here the Hg$^+$ transition is strongly dependent on $\alpha$, while the Al$^+$ clock is practically insensitive~\cite{dzuba99prl,dzuba99pra}.

The laboratory clock limits on $\dot\alpha$ may be improved by finding systems with enhanced sensitivity to $\alpha$-variation, usually denoted $q$ (see \Sec{sec:theory}). Candidates under consideration include the thorium ``nuclear clock''~\cite{peik03epl}, which would utilise the $\sim 7.5$~eV nuclear transition in $^{229}$Th~\cite{beck07prl}. Among other benefits, such a clock could have many orders-of-magnitude larger $q$-values than the optical Hg$^+$ clock transition~\cite{flambaum06prl,berengut09prl}. Optical atomic transitions with high sensitivity can be found in Yb$^+$~\cite{porsev09pra} and Th$^{3+}$~\cite{flambaum09pra}.

A different kind of sensitivity comes from transitions that have high \emph{relative} sensitivity to $\alpha$. For example, in the dysprosium atom there are two different transitions that are ``accidentally'' nearly degenerate and have $q$-values with different signs~\cite{dzuba99prl,dzuba03pra0,dzuba08pra0}. Because the levels are so close together the relative sensitivity, defined by $K = 2\Delta q/\omega$ where $\omega$ is the frequency of transitions between the two levels, is extremely high ($\sim 10^8$). The first experiment to utilise this transition gave fairly tight limits on $\alpha$-variation~\cite{cingoz07prl}, but the full enhancement was not realised because one of the levels is very broad.

In this paper we show that both kinds of sensitivity (large $q$ and large $K$) can be realized in highly-ionised atomic systems. While atomic spectroscopy in electron beam ion traps is currently not competitive with optical frequency standards (see, e.g.,~\cite{draganic03prl,crespo08cjp} and review~\cite{beiersdorfer09pscr}) the technology continues to improve, and with the enhancements in sensitivity reported here, highly-charged ions may prove to be a good system for detecting variation of $\alpha$. In \Sec{sec:theory} we show, using the Ag isoelectronic sequence as an example, why high $q$-values can occur in highly-charged ions, and how the tendency of such systems towards large transition frequencies can be overcome. In \Sec{sec:results} we show the results of our atomic calculations (\Sec{sec:method}) applied to some of the most promising ions in the sequence. In addition we identify a two-valence-electron ion, Sm$^{14+}$, which has optical transitions that are the most sensitive to potential variation of $\alpha$ ever found.

\section{Theory}
\label{sec:theory}

Using a simple analytical estimate of the relativistic effects in transition frequencies, we can see that more highly charged ions have higher sensitivity to $\alpha$ variation. Note that a ratio of frequencies (the quantity that is actually measured) does not depend on the units one uses. In this paper we use atomic units $e = m_e = \hbar = 1$ unless otherwise stated; in these units the atomic unit of energy is constant. Consider the relativistic corrections to the central-field Schr\"odinger equation for a valence electron, derived in the Pauli theory (see, e.g.~\cite{bethe57book})
\begin{align}
\label{eq:relshift1}
\Delta = &-\frac{\alpha^2}{2} \int R^2(r) \left( E_0 + V(r)\right)^2 r^2 dr \\
    &+ \frac{\alpha^2}{4} \int R(r) \left( \frac{dR}{dr} - X \frac{R}{r}\right) \frac{dV}{dr} r^2 dr \nonumber
\end{align}
which is accurate to order $(v/c)^2$. Here $R(r)$ and $E_0$ are the non-relativistic radial wavefunction and energy, respectively, $V(r)$ is the potential, and $X = j(j+1) - l(l+1)-s(s+1)$. Near the origin $V = Z/r$ (it is unscreened by core electrons) and the integrals \eref{eq:relshift1} converge as $1/r^3$. The radial $s$-wave wavefunction near the origin is given in the semiclassical approximation by (see, e.g.~\cite{sobelman72book})
\begin{equation}
R^2_s(r) \approx \frac{4 Z_a^2 Z}{\nu^3}, \qquad r \lesssim \frac{3}{2Z}
\end{equation}
where $Z_a$ is the effective charge that an external electron ``sees'' and $\nu$ is the effective principal quantum number, defined by $E_0 = -\frac{Z_a^2}{2\nu^2}$. For a single valence electron above closed shells, $Z_a = Z_i + 1$ where $Z_i$ is the ion charge, while for hydrogen-like ions $Z_a = Z$ and $\nu = n$, the principal quantum number. For higher waves $R^2_l(r)$ is proportional to the same parameter $Z_a^2 Z/\nu^3$.

Noting that the integrals converge over a distance \mbox{$r \lesssim 3/(2Z)$} and neglecting the small contribution of the $E_0$ term in \eref{eq:relshift1} (for $\nu \gg Z_a^2/Z^2$) we obtain
\begin{align}
\label{eq:relshift_vs_Za}
\Delta_n &= -\frac{Z_a^2}{2\nu^2}\frac{(Z\alpha)^2}{\nu} \\
\label{eq:relshift_vs_I}
&= -I_n \frac{(Z\alpha)^2}{\nu (j + 1/2)}
\end{align}
where $I_n$ is the ionization energy of the orbital. Actually, if we neglect the $E_0$ term, we can immediately see from~\eref{eq:relshift1} that the ratio of the relativistic corrections in our ``distorted wave'' case to the relativistic corrections in the Coulomb case is proportional to the ratio of the squared wavefunctions near the origin:
\[
\frac{\Delta_{Z_a,\nu}}{\Delta_{Z,n}}
 \propto \left. \frac{R_{Z_a,\nu}^2(r)}{R_{Z,n}^2(r)}\right|_{r \sim 0}\ .
\]
This again leads to Eqs.~\eref{eq:relshift_vs_Za} and \eref{eq:relshift_vs_I}.

In principle we could restore the $E_0$ term in \eref{eq:relshift1}. However many-body corrections are much more important than small improvements to \Eref{eq:relshift_vs_Za}. As discussed in \cite{dzuba99pra}, for single-valence electrons the relativistic energy shift can be approximately described by the equation
\begin{equation}
\label{eq:relshift_manybody}
\Delta_n = -I_n \frac{(Z\alpha)^2}{\nu} \left[ \frac{1}{j+1/2} - C(Z,j,l) \right]
\end{equation}
where $C(Z, j, l)$ depends on the atom and partial wave, but does not depend strongly on the principal quantum number.

In practice a more sophisticated numerical treatment is needed. We characterise the dependence of transition frequencies on small changes in $\alpha$ by the parameter $q$, defined by the formula
\begin{equation}
\label{eq:q-def}
\omega = \omega_0 + q\,x
\end{equation}
where $\omega_0$ is the energy at the present day value of the fine-structure constant $\alpha_0$ and
\begin{equation}
\label{eq:x-def}
x = (\alpha/\alpha_0)^2 - 1 \approx 2 \frac{\alpha - \alpha_0}{\alpha_0}\ .
\end{equation}
Our definition of $q$ reflects the fact that the relativistic shift scales as $\alpha^2$. Calculations of $\omega$ and $q$ are discussed in \Sec{sec:method}.

\begin{figure}[tb]
  \includegraphics[width=0.43\textwidth]{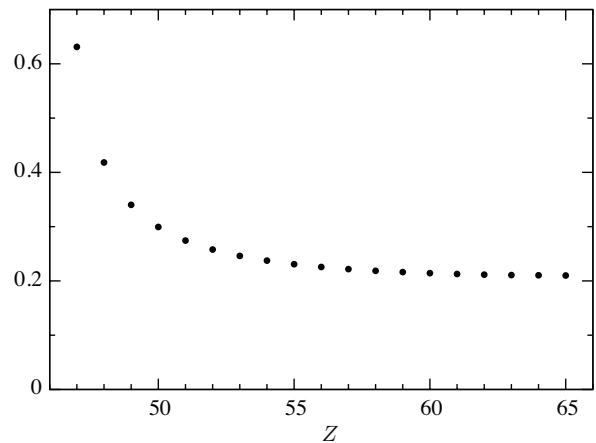}
\caption{\label{fig:q5s_norm} Ratio $-q/(Z^2\alpha^2 I_n)$ for the $5s$ level along the Ag isoelectronic sequence, calculated using the Dirac-Fock theory. It is seen that as $Z$ increases the ratio tends to a constant (which is close to $1/5$) in accordance with \eref{eq:relshift_vs_Za}.}
\end{figure}

When our formula \eref{eq:relshift_vs_I} is correct (when many-body effects are not too large), $q \approx \Delta_n/\alpha^2$. In this case, the ratio $-q/(Z^2\alpha^2 I_n)$ is a constant. In \Fig{fig:q5s_norm} we present a Dirac-Fock calculation of this ratio for different ions along the Ag isoelectronic sequence:  Ag, Cd$^+$, In$^{2+}$, etc. It is seen that as $Z$ increases the ratio indeed tends to a constant (which is close to $1/5$).

We have shown that sensitivity to $\alpha$-variation increases with ion charge as $Z_a^2 = (Z_i+1)^2$. Unfortunately, the interval between different energy levels in an ion also increases as $\sim Z_a^2$, which can quickly take the transition frequency out of the range of lasers as $Z_a$ increases.
However, the phenomena of Coulomb degeneracy and configuration crossing can be used to combat this tendency. In a neutral atom, an electron orbital with a larger angular momentum is significantly higher than one with smaller angular momentum but with the same principal quantum number $n$. On the other hand, in the hydrogen-like limit orbitals with different angular momentum but the same principal quantum number are nearly degenerate. Therefore somewhere in between there can be a crossing point where two levels with different angular momentum and principal quantum number can come close together: in such cases the excitation energy may be within laser range.

\begin{figure}[tb]
  \includegraphics[width=0.45\textwidth]{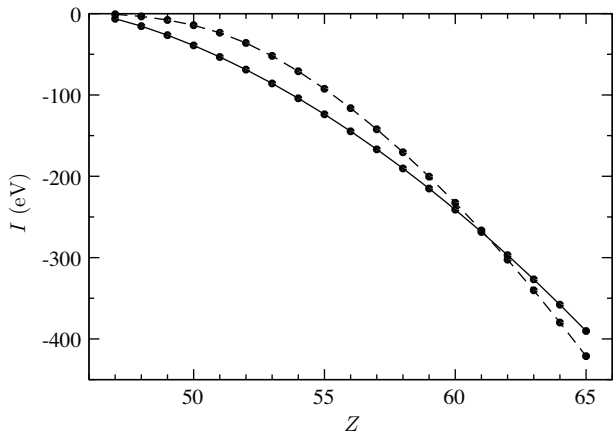}
\caption{\label{fig:AgLikeEnergy} Dirac-Fock ionisation energies of $5s$ (solid) and $4f_{7/2}$ (dashed) levels for the Ag isoelectronic sequence.}
\end{figure}

Consider again our example of the Ag isoelectronic sequence. Neutral Ag ($Z = 47$) has a single valence electron above closed shells. The ground state has the valence electron in the $5s$ orbital, while the $4f$ orbital forms an excited level.
In \Fig{fig:AgLikeEnergy} we present calculated Dirac-Fock ionisation energies of the Ag isoelectronic sequence. One can see that as $Z$ is increased, there is  a crossing point where the $4f$ level becomes the ground state. At this point, around $Z=61$, even though both levels have ionization energies of $\sim 270$~eV, the difference between them is very much smaller. In \Tref{tab:AgLike} we see that the excitation energy for ions near the crossing point is well within the range of optical lasers.

Higher-$Z$ ions with many valence electrons can have more complex behaviour as ion charge is increased. For example, in the neutral Th the $5s$ electron is a core electron while the $5f$ electron is a valence electron: therefore the $5f$ orbital is above the $5s$ orbital on the energy scale. Moreover, it even lies above the $7s$ orbital. As seen from the experimental spectrum of the energy levels of the neutral (four-valence) Th, the energy of the $6d^2 7s5f$ state is above the energy of the $6d^2 7s^2$ state. In the hydrogen-like Th the energy of the $5f$ orbital is approximately equal to the energy of the $5s$ orbital, i.e. it is significantly lower than the energy of the $7s$ state. Therefore, there should be an ion charge at which the $5f$ orbital ``crosses'' the $7s$ orbital, while at some higher charge it crosses the $6s$ orbital, etc.

\section{Method}
\label{sec:method}

To calculate energy levels we use a combination of the configuration interaction and many-body perturbation theory methods (CI+MBPT)~\cite{dzuba96pra} which has been extensively described in previous papers~\cite{berengut05pra,berengut06pra}; here we provide only a brief outline. Once we have level energies we can obtain $\alpha$-sensitivity, by repeating the calculation for $x = -0.01$ and $x=0.01$ ($x$ is defined in \eref{eq:x-def}). $q$ is then extracted from the gradient of the transition energies~\eref{eq:q-def}.

The CI+MBPT method starts with the Dirac-Fock calculation for the closed-shell core. In all cases presented here the core includes all shells up to $4spd$. In the Ag isoelectronic sequence, this corresponds to the $V^{N-1}$ approximation; for Sm$^{14+}$ it corresponds to $V^{N-2}$. A single-particle $B$-spline basis~\cite{johnson88pra} is then constructed in the potential of the nucleus and core electrons, which includes valence orbitals and a large number of excited and virtual orbitals. In this paper we use a basis $18spdf\!g$ for the CI and $30spdf\!gh$ for the MBPT.

For the two-valence-electron case, Sm$^{14+}$, we perform the full CI calculation in the frozen-core approximation. Here the many-electron wavefunction is expressed as a linear combination of Slater determinants $\left| I \right>$:
\[
\psi = \sum C_I \left| I \right>
\]
where the coefficients $C_I$ are obtained from the eigenvalue problem
\begin{equation}
\sum_{J} H_{IJ} C_J = E\,C_I
\end{equation}
and $H$ is the CI Hamiltonian. Core-valence effects are taken into account using an MBPT operator $\hat\Sigma$ which is added to the CI Hamiltonian (see Section III of \cite{berengut06pra}). We calculate $\hat\Sigma$ to second order in the residual Coulomb operator, leading to the modified eigenvalue problem
\begin{equation}
\sum_J \left( H_{IJ} + \sum_M \frac{\left< I | H | M \right>
	\left< M | H | J \right>}{E - E_M}
	\right) C_J = E\,C_I\ .
\end{equation}
The states $\left| M\right>$ include all Slater determinants of the single particle basis that have core excitations. Goldstone diagrams and analytical expressions for the one-valence-electron and two-valence-electron contributions to $\hat\Sigma$ are given in \cite{berengut06pra}. For the one-valence-electron ions presented, the CI calculation is unnecessary and we simply add the one-valence-electron part of the MBPT operator $\hat\Sigma$ to the Dirac-Fock energy.

\section{Results and discussion}
\label{sec:results}

\begin{table}[tb]
\caption{Energies and sensitivity coefficients ($q$) for isoelectronic
  sequence Nd$^{13+}$, Pm$^{14+}$ and Sm$^{15+}$  (cm$^{-1}$).}
\label{tab:AgLike}
\begin{ruledtabular}
\begin{tabular}{lccrr}
\multicolumn{1}{c}{Ion} & 
\multicolumn{1}{c}{$Z$} & 
\multicolumn{1}{c}{Level} & 
\multicolumn{1}{c}{Energy} & 
\multicolumn{1}{c}{$q$} \\
\hline
Nd$^{13+}$
 & 60 & $5s_{1/2}$ &      0 &       0  \\
 &    & $4f_{5/2}$ &  59279 &  106000  \\
 &    & $4f_{7/2}$ &  64654 &  111000  \\
Pm$^{14+}$
 & 61 & $5s_{1/2}$ &      0 &       0  \\
 &    & $4f_{5/2}$ &   3973 &  120000  \\
 &    & $4f_{7/2}$ &  10351 &  126000  \\
Sm$^{15+}$
 & 62 & $4f_{5/2}$ &      0 &       0  \\
 &    & $4f_{7/2}$ &   7480 &    7000  \\
 &    & $5s_{1/2}$ &  56272 & -136000  \\
\end{tabular}
\end{ruledtabular}
\end{table}

In \Tref{tab:AgLike} we present the results of our method for ions of the Ag isoelectronic sequence near the $5s$ -- $4f$ orbital crossing point (see \Sec{sec:theory}). It is seen that even though the ionisation energies of these levels are very large ($\sim 270$~eV), which is reflected in the very large $q$ values, the transitions themselves can be within the optical regime. High sensitivity to variation of $\alpha$ can be achieved in clocks by comparing the very sensitive $4f \rtw 5s$ transition in Sm$^{15+}$ to the $4f$ fine-structure transition, which in this case would be the anchor. Alternatively, if it is convenient to compare frequencies between ions with high accuracy, comparison of the $4f \rtw 5s$ negative-shifting transition in Sm$^{15+}$ with one of the large positive shifting $5s \rtw 4f$ transitions in Pm$^{14+}$ or Nd$^{13+}$ would allow an even higher sensitivity to $\alpha$-variation.

One of the most interesting cases is Sm$^{14+}$. It has two valence electrons above closed shells which are in the $5s$ and $4f$ states for the low-lying configurations. The states of all three configurations $4f5s$, $4f^2$ and $5s^2$ are relatively close to each other on the energy scale and are probably readily accessible to modern lasers. The transitions between these states correspond to the $4f \rtw 5s$ or $5s \rtw 4f$ single-electron transitions, which ensures strong sensitivity to the variation of the fine-structure constant. The results of calculations for Sm$^{14+}$ are presented in \Tref{tab:SmXV}. We are unaware of any experimental data for this ion, therefore only theoretical values are presented. Lifetimes are calculated using electric dipole transitions only (\Tref{tab:E1-SmXV}).

\begin{table}
\caption{Energy levels, sensitivity coefficients ($q$) 
  (cm$^{-1}$) and lifetimes ($\tau$) for lower states of Sm$^{14+}$.}
\label{tab:SmXV}
\begin{ruledtabular}
\begin{tabular}{llcrrr}
\multicolumn{2}{c}{Configuration} & $J$ &
\multicolumn{1}{c}{Energy} & 
\multicolumn{1}{c}{$q$}  & \multicolumn{1}{c}{$\tau$} \\
\hline
GS & $5s4f\ ^3\!F^o$ & 2 &      0 &       0 &  \\
   &                 & 3 &   1814 &     987 & 920~s \\
   &                 & 4 &   7282 &    6350 & 4.2~s \\
A  & $4f^2\ ^3\!H$   & 4 &    495 &  129898 &  \\
   &                 & 5 &   6288 &  135389 &  \\
   &                 & 6 &  12067 &  140213 &  \\
   & $4f^2\ ^3\!F$   & 2 &   9771 &  131140 & 16~ms \\
   &                 & 3 &  13746 &  135303 & 12~ms \\
   &                 & 4 &  14377 &  134910 & 35~ms \\
   & $5s4f\ ^1\!F^o$ & 3 &  13047 &    6819 & 50~ms \\
   & $4f^2\ ^1\!G$   & 4 &  21850 &  141015 & 11~ms \\
B  & $5s^2\ ^1\!S$   & 0 &  28248 & -124689 &  \\
\end{tabular}
\end{ruledtabular}
\end{table}

Consider, for example, the levels marked as GS (ground state), A~(495~\cm) and B~(28248~\cm). Both of these states have long lifetimes (no allowed dipole transitions) and the linewidths of transitions from the ground state to A or B are probably sufficiently narrow to ensure accuracy similar to that which is achieved in atomic clocks. If $\alpha$ varies in time, states A and B will move in opposite directions. Both of these states have some of the largest $q$-values ever seen in an atomic system.

The relative change of the ratio of the frequencies of two transitions can be written as
\begin{equation}
\label{eq:n1n2}
\frac{\Delta(\omega_1/\omega_2)}{(\omega_1/\omega_2)}
   = \left(K_1 - K_2 \right) \frac{\Delta\alpha}{\alpha}
\end{equation}
where
\begin{equation}
\label{eq:K}
K = \frac{2q}{\omega}\ .
\end{equation}
Substituting numbers from \Tref{tab:SmXV} one obtains for levels A~(495~\cm) and B~(28248~\cm):
\begin{gather}
K_A = 525, \qquad K_B = -8.8, \nonumber \\
\frac{\Delta(\omega_A/\omega_B)}{(\omega_A/\omega_B)}
   = 534 \frac{\Delta\alpha}{\alpha}\ .
\end{gather}
This is more than two orders-of-magnitude higher relative sensitivity than that of the Hg$^+$ frequency standard ($K = -3.19$), the system in which the current strongest constraint on the present-day time variation of $\alpha$ was obtained~\cite{rosenband08sci}. The very high sensitivity $K_A$ comes from large $q$ and small $\omega$, and we should note that our theoretical transition frequencies result from a cancelation of energy levels at the 0.01 -- 0.1\% level, and may only be accurate to perhaps $\sim 2000~\cm$: enough to reduce the relative sensitivity $K_A$ considerably. On the other hand the $q$-values in \Tref{tab:SmXV} are very stable in our calculation. Comparison of any of the lines in \Tref{tab:SmXV} with B will give values of $q_1 - q_2 \approx 260\,000~\cm$. This represents an absolute enhancement in $\alpha$-sensitivity five times that of the Hg$^+$/Al$^+$ clock comparison.

\begin{table}
\caption{Reduced electric-dipole matrix elements for transitions between lower states of Sm$^{14+}$ (a.u.)}
\label{tab:E1-SmXV}
\begin{ruledtabular}
\begin{tabular}{rcrcc}
\multicolumn{2}{c}{Level 1} &
\multicolumn{2}{c}{Level 2} & Amplitude \\
\multicolumn{1}{c}{Configuration} & $J$ & 
\multicolumn{1}{c}{Configuration} & $J$ & 
\multicolumn{1}{c}{$\langle 2||d||1\rangle$}  \\
\hline
$4f^2\ ^3\!F$ & 2 & $5s4f\ ^3\!F^o$ & 2 & 0.011800 \\
              &   & $5s4f\ ^3\!F^o$ & 3 & 0.007480 \\
$4f^2\ ^3\!F$ & 3 & $5s4f\ ^3\!F^o$ & 2 & 0.002030 \\
              &   & $5s4f\ ^3\!F^o$ & 3 & 0.012300 \\
              &   & $5s4f\ ^3\!F^o$ & 4 & 0.006650 \\
              &   & $5s4f\ ^1\!F^o$ & 3 & 0.005530 \\
$4f^2\ ^3\!F$ & 4 & $5s4f\ ^3\!F^o$ & 3 & 0.004450 \\
              &   & $5s4f\ ^3\!F^o$ & 4 & 0.015500 \\
              &   & $5s4f\ ^1\!F^o$ & 3 & 0.017900 \\
$4f^2\ ^1\!G$ & 4 & $5s4f\ ^3\!F^o$ & 3 & 0.001850 \\
              &   & $5s4f\ ^3\!F^o$ & 4 & 0.006820 \\
              &   & $5s4f\ ^1\!F^o$ & 3 & 0.018100 \\

$5s4f\ ^3\!F^o$ & 3 & $4f^2\ ^3\!H$ & 4 & 0.001280 \\
$5s4f\ ^1\!F^o$ & 3 & $4f^2\ ^3\!H$ & 4 & 0.005890 \\
                &   & $4f^2\ ^3\!F$ & 2 & 0.001970 \\
$5s4f\ ^3\!F^o$ & 4 & $4f^2\ ^3\!H$ & 4 & 0.001840 \\
                &   & $4f^2\ ^3\!H$ & 5 & 0.000078 \\
\end{tabular}
\end{ruledtabular}
\end{table}


The lifetimes depend more strongly on the transition frequencies than on the dipole amplitudes. On the other hand, the accuracy of the transition frequencies may be low since they are obtained as the difference of the two-electron removal energies of two levels. Therefore, along with the lifetimes of the levels, we also present the calculated values of the transition amplitudes (\Tref{tab:E1-SmXV}). If frequencies are measured then the lifetimes can be recalculated using these amplitudes. Note that the amplitudes are small. This is because they correspond to $s$--$f$ single-electron transitions, which cannot be an electric dipole transition. Therefore, the E1 transition amplitudes in the two-electron states are due to configuration mixing with suitable states, and this mixing is small.

\section{Conclusion}

We have shown that highly-charged ions present opportunities for optical transitions with very high sensitivity to $\alpha$-variation. The enhancement is proportional to the ionization energy of the states, which increases with ion charge as $\sim (Z_i+1)^2$. We have presented a method to identify good experimental candidate ions that have suitable transitions within the range of optical lasers. Applying our method to the Ag isoelectronic sequence allowed us to identify Sm$^{14+}$ as having suitable transitions with the highest $q$-values seen to date in an atomic system. Furthermore it has a very high relative sensitivity, and is therefore an excellent candidate for studies of temporal $\alpha$-variation.

The ideas presented in this letter may be extended to other isoelectronic sequences, Au, Hg, etc., which would benefit from a larger $Z^2$ enhancement, along with the $(Z_i + 1)^2$ enhancement we have discussed.

\acknowledgments

We thank J. R. Crespo L\'opez-Urrutia for useful discussions. This work was supported in part by the Australian Research Council. We thank the NCI National Facility for valuable computer time.

\bibliography{references}

\end{document}